\documentclass[a4paper,reqno]{amsart}

\usepackage{amsmath}
\usepackage{amssymb}

\usepackage{hyperref}

\newcommand{\cB}{\mathcal{B}}
\newcommand{\cD}{\mathcal{D}}
\newcommand{\cE}{\mathcal{E}}
\newcommand{\cM}{\mathcal{M}}
\newcommand{\cQ}{\mathcal{Q}}

\newcommand{\C}{\mathbb{C}}
\newcommand{\I}{\mathbb{I}}
\newcommand{\R}{\mathbb{R}}

\renewcommand{\vec}[1]{\mathbf{#1}}

\newcommand{\then}{\Rightarrow}
\newcommand{\norm}[1]{\|#1\|}
\newcommand{\propersubset}{\varsubsetneq}

\newcommand{\secref}[1]{\textsc{Section \ref{#1}}}
\newcommand{\defref}[1]{\textbf{Definition \ref{#1}}}

\newtheorem{definition}{Definition}

\DeclareMathOperator{\Tr}{Tr}
\DeclareMathOperator{\im}{i}
\DeclareMathOperator{\Span}{Span}

\begin{document}
\title{Probing the geometry of quantum states with symmetric POVMs}
\author{Jos\'e Ignacio Rosado}

\email{joseirs@hotmail.com}


\date{\today}

\begin{abstract}
    The geometry of the Quantum State Space, described by Bloch vectors, is a very intricate one. A deeper understanding of this geometry could lead to the solution of some difficult problems in Quantum Foundations and Quantum Information such as the existence of SIC-POVMs and the cardinality of the maximal set of MUBs. In this paper we show that the geometry of quantum states can be described by the probability distributions that quantum states induce over the outcomes of symmetric POVMs not necessarily of arbitrary rank or informationally complete. We also describe the properties of these symmetric POVMs.
\end{abstract}

\maketitle

\section{Introduction}
SIC-POVMs, Symmetric and informationally complete POVMs with all its elements of rank one, have been studied extensively due to their importance in Quantum Information and in Quantum Foundations as the more efficient measurements to gain information about quantum states. However the lack of a proof of their existence in any dimension and the difficulty to construct them led D. Appleby \cite{Appleby07} to the introduction of symmetric and informationally complete POVMs of arbitrary rank which he called SI-POVMs, in the hope that they might lead to demonstrate the existence of SIC-POVMs in every finite dimension. Although this hope has not been fulfilled SI-POVMs have their own interest and deserve to be studied. In this paper, \secref{s:Symmetric POVMs}, we will study not only the symmetric POVMs introduced by Appleby but also symmetric POVMs with $2,3,\ldots$ or $d^2$ outcomes ($d$ is the dimension of the Hilbert space associated to the physical system of interest). Recent publications studying general symmetric and informationally complete POVMs are \cite{Gour13} where it is proved that SI-POVMs exist in all dimensions and provides a method for constructing them and \cite{Rastegin13} where some statistical properties of SI-POVMs are studied. Our interest on symmetric POVMs will be its utility in the study of geometry of quantum states.

In $d$-dimensional quantum mechanics the set of states, consisting of hermitian, positive and normalized density matrices, has $d^2-1$ real dimensions and can be represented as a subset, $\cB^{(d)}$, of $\R^{d^2-1}$ by means of the Bloch vectors, \secref{s:Parameterization of density matrices}. For $d=2$ the set $\cB^{(2)}$ is the Bloch ball that represents for example the polarization states of a single photon. For $d>2$ the geometry of $\cB^{(d)}$ is very intricate \cite{Bengtsson&Weis&Zyczkowski13}. Given a symmetric POVM, $\cE$, of $N$ outcomes we construct the set $\cQ_{N-1}^{(d)}\subset\R^{d^2-1}$ (the subindex indicates the dimension of the set) of probability distributions over the outcomes of $\cE$ that correspond to quantum states, \secref{s:Representation of quantum states as probability distributions}. We will see that $\cQ_{N-1}^{(d)}$ is the orthogonal projection of $\cB^{(d)}$ onto some $(N-1)$-dimensional subspace of $\R^{d^2-1}$, and that the orientation of $\cQ_{N-1}^{(d)}$ can be any one we want, so with the statistics of different symmetric POVMs we can explore the geometry of $\cB^{(d)}$. A special case is $N=d^2$ then $\cE$ is informationally complete and $\cQ^{(d)}$ is exactly equal to the set $\cB^{(d)}$ up to a scale factor. The geometry of $\cB^{(d)}$ was described in \cite{Kimura03} in terms of the coefficients of the characteristic polynomial of density matrices, a refinement of this description \cite{Kimura&Kossakowski05} was introduced, this improvement is what the authors called the spherical-coordinate point of view which will be useful in the present paper.

\section{Parameterization of density matrices}\label{s:Parameterization of density matrices}
The set of density matrices associated to a $d$-dimensional quantum system will be denoted by $\cD^{(d)}$, then
\begin{equation}\label{e:DefD(d)}
  \cD^{(d)}=\{\rho\in \cM_{d\times d}(\C)\quad :\quad \Tr(\rho)=1,\,\,\rho^{\dag}=\rho,\,\,\rho\geq0\},
\end{equation}
where $\cM_{d\times d}(\C)$ is the set of all complex matrices of order $d$.

Any element of $\cD^{(d)}$ can be parameterized as
\begin{equation}\label{e:rho(b)}
  \rho(\vec b)=\frac{1}{d}\I_d+\vec b\cdot\boldsymbol\sigma,
\end{equation}
where $\I_d$ is the identity matrix in $d$ dimensions, $\vec b$ is a vector in $\R^{d^2-1}$ and $\boldsymbol\sigma=(\sigma_1,\sigma_2,\ldots,\sigma_{d^2-1})$ are traceless Hermitian matrices which form a basis of the algebra $\mathfrak{su}(d)$. Such basis of $\mathfrak{su}(d)$ is chosen so that \cite[p.~422]{Bengtsson&Zyczkowski06}
\begin{align}
    [\sigma_a,\sigma_b] &=2\im f_{abc}\sigma_c\,,\label{e:sigmacommutator}\\
    \{\sigma_a,\sigma_b\} &=\frac{4}{d}\:\delta_{ab}+2\:d_{abc}\sigma_c\,,\label{e:sigmaanticommutator}\\
    \Tr(\sigma_a\sigma_b) &=2\:\delta_{ab}\,,\label{e:trsigmasigma}\\
    \Tr(\sigma_a\sigma_b\sigma_c) &=2\:d_{abc}+2\im f_{abc}\label{e:trsigmasigmasigma},
\end{align}
where $d_{abc}\in\R$ is totally symmetric, $f_{abc}\in\R$ is totally antisymmetric, the indices $a$, $b$, $c$ are in the set $\{1,2,\ldots,d^2-1\}$ and we follow on them the Einstein summation convention.

Not all vectors $\vec b\in\R^{d^2-1}$ substituted in \eqref{e:rho(b)} give rise to an element of $\cD^{(d)}$, the ones such that $\rho(\vec b)\in\cD^{(d)}$ will be called Bloch vectors and the set of all these vectors will be denoted by $\cB^{(d)}$, so
\begin{equation}\label{e:DefB(d)}
  \cB^{(d)}=\{\vec b\in\R^{d^2-1}\quad :\quad \rho(\vec b)\in\cD^{(d)}\},
\end{equation}
where $\rho(\vec b)$ is given by \eqref{e:rho(b)}. With this definition \eqref{e:rho(b)} establishes a bijection between the sets $\cB^{(d)}$ and $\cD^{(d)}$ therefore the state of a quantum $d$-dimensional system can be represented either by a density matrix, $\rho\in\cD^{(d)}\subset\cM_{d\times d}(\C)$, or by a Bloch vector $\vec b\in\cB^{(d)}\subset\R^{d^2-1}$.

Two very important parameters characterising $\cB^{(d)}$ are its \emph{outradius}, $R_{out}^{(d)}$, the radius of the smallest circumscribed sphere, and its \emph{inradius}, $r_{in}^{(d)}$, the radius of the largest inscribed sphere, their values are \cite[p.~220]{Bengtsson&Zyczkowski06}
\begin{align}
  R_{out}^{(d)} &= \sqrt{\frac{d-1}{2d}} && \text{and}\label{e:outradius}\\
  r_{in}^{(d)} &= \frac{1}{\sqrt{2d(d-1)}}. \label{e:inradius}
\end{align}
Then all Bloch vectors corresponding to pure states, and only these, lie on the outer sphere but the converse is, in general, not true, not all points on the outer sphere correspond to Bloch vectors. In fact given $\vec b\in\R^{d^2-1}$ a point on the outer sphere, ie, $\norm{\vec b}=\sqrt{(d-1)/(2d)}$ it can be proved \cite[p.~215]{Bengtsson&Zyczkowski06} that
\begin{equation}\label{e:Conditions b pure}
  \vec b\in\cB^{(d)}\,\,\iff\,\,\vec b \star\vec b=\frac{d-2}{d}\vec b\,,
\end{equation}
where $(\vec b \star\vec b)_a=d_{abc}b_b\,b_c$ (we sum over repeated indices) and $d_{abc}$ are the structure constants that appear in \eqref{e:sigmaanticommutator}. For $d=2$ all the structure constants $d_{abc}$ are identically zero and so the right condition in \eqref{e:Conditions b pure} is a trivial identity, therefore in this case all points on the outer sphere correspond to pure states. For $d>2$ the conditions $\vec b \star\vec b=(d-2)/d\;\vec b$ defines a $(2d-2)$-dimensional submanifold on the outer sphere very difficult to apprehend. Such difficulty underlies, I think, the lack of answers to some problems in Quantum Foundations, for example the problem on the existence of SIC-POVMs or the the lack of a proof that $d+1$ mutually unbiased basis in $d$-dimensional Hilbert space exist only when $d$ is the power of a prime.

For the inner ball we have, by definition,
\begin{equation}\label{e:Condition b inner}
  \norm{\vec b}\leq r_{in}^{(d)}\; \then\; \vec b\in\cB^{(d)}.
\end{equation}
So when $r_{in}^{(d)}<\norm{\vec b}\leq R_{out}^{(d)}$ only some directions are available for $\vec b\in\cB^{(d)}$, but when $\norm{\vec b}\leq r_{in}^{(d)}$ all directions are possible for $\vec b\in\cB^{(d)}$.

These relations between norm and available directions of the Bloch vectors are the reason we use hereinafter the spherical-coordinate point of view introduced by G. Kimura and A. Kossakowski in \cite{Kimura&Kossakowski05}. This point of view consists in writing each Bloch vector as the product of two factors: a scalar and a vector of constant norm
\begin{equation}\label{e:b(kappa,n)}
  \vec b=\kappa\vec n,
\end{equation}
our election for these two factors is as follows
\begin{align}
  \kappa &\in [0,1]\label{e:rango kappa}, \\
  \norm{\vec n} &= \sqrt{\frac{d-1}{2d}} \label{e:norm n}.
\end{align}
It is obvious from \eqref{e:outradius} that any Bloch vector can be written as in \eqref{e:b(kappa,n)} with the choices \eqref{e:rango kappa} and \eqref{e:norm n}. Substituting \eqref{e:b(kappa,n)} in \eqref{e:rho(b)} we obtain a new parameterization of density matrices
\begin{equation}\label{e:rho(kappa,n)}
  \rho(\kappa,\vec n)=\frac{1}{d}\I_d+\kappa\vec n\cdot\boldsymbol\sigma,
\end{equation}
in this way we see that $\kappa$ is a kind of \emph{purity index}, with $\kappa=0$ for the maximally mixed state and $\kappa=1$ for pure states, on the other hand $\vec n$, having its norm fixed, is a \emph{directional vector}. The existence of an inner sphere with the maximal radius \eqref{e:inradius} in $\cB^{(d)}$ means that for any dimension $d$
\begin{equation}\label{e:KappaInner}
  \kappa\in\left[0,\frac{1}{d-1}\right]\implies\rho(\kappa,\vec n)\in\cD^{(d)},
\end{equation}
independently of the direction of $\vec n$.

In what follows we will need sometimes the trace of the product of two density matrices, so we write it now
\begin{align}
  \Tr(\rho(\kappa_1,\vec n_1)\rho(\kappa_2,\vec n_2)) &= \frac{1}{d}+2 \kappa_1\kappa_2 \vec n_1\cdot\vec n_2\label{e:ProdRhos(n1,n2)}\\
  &= \frac{1+(d-1)\kappa_1\kappa_2\cos\theta_{12}}{d}.\label{e:ProdRhos}
\end{align}
As can be deduced from the tracelessness of $\mathfrak{su}(d)$ generators and equations \eqref{e:trsigmasigma} and \eqref{e:norm n}. We have also introduced $\theta_{12}$, it is the angle between vectors $\vec n_1$ and $\vec n_2$. This trace must be nonnegative then
\begin{equation}\label{e:lower bound kappa1 kappa2 cos(theta)}
  \kappa_1\kappa_2\cos\theta_{12}\geq -\frac{1}{d-1},
\end{equation}

It is interesting to see that when both states have the same purity $\kappa$ then the angle between the corresponding directional vectors can have an upper bound, other than $\pi$. First we write \eqref{e:lower bound kappa1 kappa2 cos(theta)} with $\kappa_1=\kappa_2=\kappa$
\begin{equation}\label{e:low b cos(theta)}
  \cos\theta_{12}\geq -\frac{1}{\kappa^2(d-1)}.
\end{equation}
If we want the right hand side of \eqref{e:low b cos(theta)} to be a nontrivial lower bound for $\cos\theta_{12}$ then
\begin{equation*}
  -\frac{1}{\kappa^2(d-1)} > -1
\end{equation*}
and therefore
\begin{equation}\label{e:kappa with theta bounded}
  \kappa > \frac{1}{\sqrt{d-1}}.
\end{equation}
Then defining the angles
\begin{equation}\label{e:theta(kappa)}
  \Theta_{\kappa}=
  \begin{cases}
    \pi & \text{if $0< \kappa\leq\frac{1}{\sqrt{d-1}}$}\\
    \arccos\left[-\frac{1}{\kappa^2(d-1)}\right] & \text{if $\frac{1}{\sqrt{d-1}}<\kappa\leq 1$}
  \end{cases}
\end{equation}
and from \eqref{e:low b cos(theta)} we can conclude that Bloch vectors with purity $\kappa$ cannot be separated by an angle greater than $\Theta_{\kappa}$.

\section{Symmetric POVMs}\label{s:Symmetric POVMs}
First we introduce the concept of a symmetric positive operator valued measure (symmetric POVM) of arbitrary rank \cite{Appleby07}, and of arbitrary number of outcomes.

\begin{definition}\label{d:symmetric POVM}
  A measurement of a $d$-dimensional quantum system is a symmetric POVM with $N$ outcomes if and only if to each outcome we can associate an element of a set, $\cE=\{E_i\}_{i=1}^N$, of positive semi-definite operators such that for all $i,j\in\{1,2,\ldots,N\}$ the following conditions are satisfied
  \begin{align}
    \sum_{i=1}^N E_i &= \I_d\,,\label{e:EiCompleteness}\\
    \Tr(E_iE_j) &= \alpha+\beta\delta_{ij} && \text{and}\label{e:TrEiEj}\\
    i\neq j &\implies E_i\neq E_j\,. \label{e:Eis diferentes}
  \end{align}
\end{definition}
The condition \eqref{e:EiCompleteness} implies that for any state we always obtain an outcome of our symmetric POVM if we make the corresponding measurement, it is the completeness condition. The symmetry of the POVM is encoded in condition \eqref{e:TrEiEj}. Finally condition \eqref{e:Eis diferentes} means that different outcomes provide different information about the quantum system, it is also necessary to avoid the uninteresting case in which all operators are proportional to the identity.

In \defref{d:symmetric POVM} the parameters $\alpha$ and $\beta$ are not independent. If we sum over the index $j$ in \eqref{e:TrEiEj} we obtain
\begin{equation}\label{e:Ei(alfa,beta)}
  \Tr(E_i) = N\alpha+\beta,
\end{equation}
where we have used \eqref{e:EiCompleteness}. Now we sum over the index $i$ of \eqref{e:Ei(alfa,beta)}
\begin{align*}
  \Tr\left(\sum_{i=1}^N E_i\right) &= \sum_{i=1}^N (N\alpha+\beta)\\
  \Tr(\I_d) &= N(N\alpha+\beta) && \text{by \eqref{e:EiCompleteness}}\\
  d &= N(N\alpha+\beta).
\end{align*}
Therefore
\begin{equation}\label{e:alfa(beta)}
  \alpha=\left(\frac{d}{N}-\beta\right)\frac{1}{N}.
\end{equation}
Substituting \eqref{e:alfa(beta)} in \eqref{e:Ei(alfa,beta)} we obtain
\begin{equation}\label{e:trEi}
  \Tr(E_i) = \frac{d}{N}
\end{equation}
and the same substitution in \eqref{e:TrEiEj} results in
\begin{equation}\label{e:TrEiEj(beta)}
  \Tr(E_iE_j) = \frac{1}{N}\left(\frac{d}{N}+(N\delta_{ij}-1)\beta\right)
\end{equation}

Since each $E_i$ is a positive operator then $E_i/\Tr E_i$ is a density matrix $\rho_i$, therefore using \eqref{e:trEi} and \eqref{e:rho(kappa,n)} we can write each element of $\cE$ as
\begin{align}
  E_i &= \frac{d}{N}\,\rho_i = \frac{d}{N}\,\rho(\kappa_i,\vec n_i)\label{e:Ei(rho)}\\
  &= \frac{d}{N}\left(\frac{1}{d}\I_d+\kappa_i\vec n_i\cdot\boldsymbol\sigma\right)\quad i\in\{1,2,\ldots,N\} \label{e:Ei(kappa,n)}.
\end{align}

Now we will find what conditions $\kappa_i$ and $\vec n_i$ must satisfy in order that $\cE=\{E_i\}_{i=1}^N$ is a symmetric POVM according to \defref{d:symmetric POVM}. We begin calculating $\Tr(E_iE_j)$
\begin{align}
  \Tr(E_iE_j) &= \left(\frac{d}{N}\right)^2\Tr[\rho(\kappa_i,\vec n_i)\rho(\kappa_j,\vec n_j)]\notag\\
  &= \left(\frac{d}{N}\right)^2\frac{1+(d-1)\kappa_i\kappa_j\cos\theta_{ij}}{d} && \text{by \eqref{e:ProdRhos}}\notag\\
  &=\frac{d}{N}\,\frac{1+(d-1)\kappa_i\kappa_j\cos\theta_{ij}}{N}.\label{e:TrEiEj(kappa,d)}
\end{align}
Now we equate this expression with \eqref{e:TrEiEj(beta)} so that we have a symmetric POVM. For $i=j$ in both equations, we obtain
\begin{equation*}
  \frac{d}{N}\, \frac{1+(d-1)\kappa_i^2}{N} = \frac{1}{N}\left(\frac{d}{N}+(N-1)\beta\right)
\end{equation*}
then
\begin{equation}\label{e:beta(kappa)}
  \beta=\frac{d(d-1)}{N(N-1)}\,\kappa_i^2.
\end{equation}
This equation holds for all $i\in\{1,2,\ldots,N\}$, so all $\kappa_i$ have the same value, we will call this value $\kappa$. The value $\kappa=0$ is not very interesting because in order to satisfy \defref{d:symmetric POVM} the symmetric POVM would be $\cE=\{\I_d\}$, in what follows we will suppose that $\kappa\in (0,1]$ and therefore $N>1$.

Now equating \eqref{e:TrEiEj(beta)} and \eqref{e:TrEiEj(kappa,d)} for $i\neq j$ and taking into account \eqref{e:beta(kappa)} we obtain
\begin{equation}\label{e:CosThetaij}
  \cos\theta_{ij}=-\frac{1}{N-1}.
\end{equation}
As we see for $i\neq j$ the angle between the corresponding directional vectors only depends on $N$, the number of outcomes of our symmeric POVM, and not on its particular orientation. In addition \eqref{e:CosThetaij} means that the directional vectors are the vertices of an $(N-1)$-dimensional regular simplex centered at the origin of $\R^{d^2-1}$ so we will denote the angle that appears in \eqref{e:CosThetaij} by $\theta_{N-1}$. An immediate consequence is that $N$ can only take the values $2,\;3\,\ldots$ or $d^2$ because in $\R^{d^2-1}$ we cannot have a simplex with more than $d^2$ vertices. But in fact $N$ is more restricted for some values of $\kappa$. Indeed, on one hand we have that any two directional vectors of a symmetric POVM with $N$ outcomes have an angle $\theta_{N-1}$, on the other hand we saw in \eqref{e:theta(kappa)} that the angle between Bloch vectors with purity $\kappa$ cannot be grater than $\Theta_{\kappa}$, therefore if the symmetric POVM has parameter $\kappa$ in \eqref{e:Ei(kappa,n)} then we must have $\theta_{N-1}\leq\Theta_{\kappa}$ this means that
\begin{align*}
  \cos\theta_{N-1} &\geq \cos\Theta_{\kappa}\\
  -\frac{1}{N-1} &\geq
  \begin{cases}
    -1 & \text{if $0< \kappa\leq\frac{1}{\sqrt{d-1}}$},\\
    -\frac{1}{\kappa^2(d-1)} & \text{if $\frac{1}{\sqrt{d-1}}<\kappa\leq 1$}
  \end{cases}
  && \text{by \eqref{e:theta(kappa)} and \eqref{e:CosThetaij}.}
\end{align*}
Thus depending on $\kappa$ we define
\begin{equation}\label{e:Nmin}
  N_{min}(\kappa)=
  \begin{cases}
    2 & \text{if $0< \kappa\leq\frac{1}{\sqrt{d-1}}$},\\
    \lceil\kappa^2(d-1)\rceil+1 & \text{if $\frac{1}{\sqrt{d-1}}<\kappa\leq 1$}
  \end{cases},
\end{equation}
where $\lceil x\rceil$, with $x\in\R$, is the smallest integer greater than or equal to $x$. We have proved that the number of outcomes for symmetric POVM with parameter $\kappa$ cannot be less than $N_{min}(\kappa)$. For example a symmetric POVM with $\kappa=1$, its elements are subnormalized pure states, cannot have less than $d$ outcomes.

Finally we substitute \eqref{e:CosThetaij} in \eqref{e:TrEiEj(kappa,d)} to write $\Tr(E_iE_j)$ as a function only of $N,\;d$ and $\kappa$. First when $i=j$
\begin{equation}\label{e:TrE^2}
  \Tr(E^2_i)=\frac{d}{N}\,\frac{1+(d-1)\kappa^2}{N},
\end{equation}
and now for $i\neq j$ 
\begin{equation}\label{e:TrEiEjFinal}
  \Tr(E_iE_j)=\frac{d}{N^2}\,\left(1-\frac{d-1}{N-1}\kappa^2\right).
\end{equation}
With these expressions it is easy to see that if $\kappa=1$ and $N=d$ the symmetric POVM is an ordinary von Neumann measurement.

In this section we have learned to construct a symmetric POVM,
\begin{align}\label{e:symmPOVM notation}
  \cE\left(\kappa,\{\vec n_i\}_{i=1}^N\right) &= \{E_i\}_{i=1}^N \notag\\
  &=\left\{\frac{d}{N}\rho(\kappa, \vec n_i)\right\}_{i=1}^N
\end{align}
with $N\in\{2,\,3,\,\ldots,\,d^2\}$ outcomes for a $d$-dimensional quantum system. First we take a set of vectors $\{\vec n_i\}_{i=1}^N$, all of them with norm \eqref{e:norm n}, pointing towards the vertices of an $(N-1)$-dimensional and regular simplex centered at the origin in $\R^{d^2-1}$. The final and critical step is to adjust $\kappa$ so that each $E_i$ in \eqref{e:Ei(kappa,n)} is a nonnegative operator or equivalently $\kappa\vec n_i$ is a Bloch vector. But by \eqref{e:KappaInner} this is always possible, irrespective of the orientation of the simplex defined by the directional vectors associated to the symmetric POVM at least for $0<\kappa\leq 1/(d-1)$ because then we are in the inner ball of $\cB^{(d)}$. If we want a symmetric POVM with $1/(d-1)<\kappa\leq 1$ the orientation of the simplex defined by $\{\vec n_i\}_{i=1}^N$ is in general not arbitrary and to find such a regular simplex in $\cB^{(d)}$ can be a very difficult problem.

\section{Representation of quantum states as probability distributions}\label{s:Representation of quantum states as probability distributions}
In this section we will see how to represent a quantum state as a probability distribution over the outcomes of a symmetric POVM. Probability distributions, over $N$ outcomes, can be represented by points in an $(N-1)$-dimensional regular simplex \cite[section~1.5]{Bengtsson&Zyczkowski06}. This simplex is usually embedded in $\R^{N-1}$, however we will embed it in $\R^{d^2-1}$, for all $N\in\{2,\,3,\,\ldots,\,d^2-1\}$, remember that $d$ is the dimension of the Hilbert space associated to our quantum system. The probability simplex is then a subset, $\Delta_{N-1}^{(d)}$, of $\R^{d^2-1}$ defined as follows
\begin{equation}\label{e:DefProbSimplex}
  \Delta_{N-1}^{(d)}=\left\{\vec v=\sum_{i=1}^N p_i\vec t_i\; :\; p_i\geq 0,\,i=1,\,2,\,\ldots,\,N\;\text{and}\;\sum_{i=1}^N p_i=1\right\},
\end{equation}
where all the vectors $\vec t_i\in\R^{d^2-1}$ have the same norm and the angle between any two of them is $\theta_{N-1}$. Given $\vec v\in\Delta_{N-1}^{(d)}$ its barycentric coordinates with respect to the vertices of the probability simplex are  a probability distribution over $N$ outcomes, viceversa, any probability distribution over $N$ outcomes gives rise to a vector in $\Delta_{N-1}^{(d)}$, so we will refer to the points of $\Delta_{N-1}^{(d)}$ indifferently as vectors or as probability distributions.

Given a quantum state
\begin{equation}\label{e:GenState}
  \rho=\frac{1}{d}\I_d+\vec b\cdot\boldsymbol\sigma\in\cD^{(d)},
\end{equation}
(this time we don't factorize the Bloch vector, $\vec b$, into a purity index and a directional vector) we want to find the probability distribution induced by $\rho$ over the outcomes of the symmetric POVM $\cE\left(\kappa,\{\vec n_i\}_{i=1}^N\right)$, with $N\geq 2$. The probability of obtaining outcome $i\in\{1,\,2,\,\ldots,\,N\}$, in the measurement just described is
\begin{align}
  p_i &= \Tr(\rho E_i)\notag\\
  &= \frac{d}{N}\Tr(\rho\rho_i) && \text{by \eqref{e:Ei(rho)}}\notag\\
  &= \frac{d}{N}\left(\frac{1}{d}+2\kappa \vec b\cdot\vec n_i\right) && \text{by \eqref{e:ProdRhos(n1,n2)}}.\label{e:pi(kappa,b,ni)}
\end{align}
Now we are going to represent this probability distribution as a vector in a probability simplex \eqref{e:DefProbSimplex}, to this end we need to choose the vectors $\{\vec t_i\}_{i=1}^N$, in principle, $\Delta_{N-1}^{(d)}$ can have any orientation and scale but as we already have a regular $(N-1)$-dimensional simplex, the one defined by the directional vectors $\{\vec n_i\}_{i=1}^N$ of the symmetric POVM $\cE\left(\kappa,\{\vec n_i\}_{i=1}^N\right)$, it is natural to use them to define the probability simplex. We will choose the scale of the probability simplex $\Delta_{N-1}^{(d)}$ to be a convenient one, namely
\begin{equation}\label{e:t(n)}
  \vec t_i=\frac{N-1}{d-1}\,\vec n_i\quad \forall\, i\in\{1,\,2,\,\ldots,\,N\}.
\end{equation}
The convenience of this scale will be seen below.

The points of $\Delta_{N-1}^{(d)}$ corresponding to probability distributions associated to quantum states will be denoted by $\cQ_{N-1}^{(d)}$. It is important to note that in general $\cQ_{N-1}^{(d)}\propersubset\Delta_{N-1}^{(d)}$.

The vector in $\Delta_{N-1}^{(d)}$ associated to $\rho$ is then
\begin{align}
  \vec v &= \sum_{i=1}^N p_i\vec t_i\label{e:def v}\\
  &= \frac{d}{N}\frac{N-1}{d-1}\sum_{i=1}^N \left(\frac{1}{d}+2\kappa \vec b\cdot\vec n_i\right)\vec n_i, && \text{by \eqref{e:pi(kappa,b,ni)}and \eqref{e:t(n)}}\notag
\end{align}
and because $\sum_{i=1}^N\vec n_i=\vec 0$, we have
\begin{equation}\label{e:v(b,n)}
  \vec v = 2\kappa\frac{d}{N}\frac{N-1}{d-1}\sum_{i=1}^N (\vec b\cdot\vec n_i)\vec n_i.
\end{equation}
To see more clearly the relation between $\vec v$ and $\vec b$, let us multiply both sides of \eqref{e:v(b,n)} by $\vec n_j$. Then using that the norm of $\vec n_i$ is \eqref{e:norm n} and that the angle between any two of them is $\theta_{N-1}$, so its cosine is \eqref{e:CosThetaij}, we obtain
\begin{equation}\label{e:vnj=kappabnj}
  \vec v\cdot\vec n_j=\kappa \vec b\cdot\vec n_j \quad\forall j\in\{1,\,2,\,\ldots,\,N\},
\end{equation}
from where we deduce that
\begin{equation}\label{e:v(b)}
  \vec v=\kappa \vec b^{\|},
\end{equation}
where $\vec b^{\|}$ is the orthogonal projection of $\vec b$ onto $\Span\left(\{\vec n_i\}_{i=1}^N\right)$. As \eqref{e:v(b)} is valid for any vector $\vec b\in\cB^{(d)}$ it follows that
\begin{equation}\label{e:Q(N-1)=Bparall}
  \cQ_{N-1}^{(d)}=\kappa\cB^{(d)\|}.
\end{equation}
So the set of quantum probability distributions over the outcomes of a symmetric POVM, $\cE\left(\kappa,\{\vec n_i\}_{i=1}^N\right)$, gives the orthogonal projection of $\cB^{(d)}$ onto the subspace generated by $\{\vec n_i\}_{i=1}^N$. As $\kappa$ decreases the set $\cQ_{N-1}^{(d)}\subset\Delta_{N-1}^{(d)}$ becomes smaller and so the distance between different probability distributions, then it is harder to distinguish between different states, in fact $\kappa$ is directly related with the efficiency of the measurement \cite{Appleby07}.

An important case happens when our symmetric POVM, $\cE$, has $N=d^2$ then $\Span\left(\{\vec n_i\}_{i=1}^N\right)=\R^{d^2-1}$ and so $\vec b^{\|}=\vec b$, then in this case the probability distribution over the outcomes of $\cE$ is sufficient to determine any Bloch vector $\vec b$, and therefore any quantum state $\rho$. Such symmetric POVMs are called informationally complete. Therefore if $\cE$ is a symmetric and informationally complete POVM we have
\begin{equation}\label{e:v(b)SI-POVMS}
  \vec v=\kappa \vec b,
\end{equation}
for any state \eqref{e:GenState} or what is the same thing
\begin{equation}\label{e:Q=B}
  \cQ^{(d)}=\kappa\cB^{(d)},
\end{equation}
where $\cQ^{(d)}\equiv \cQ_{d^2-1}^{(d)}$. In \cite{Rosado11} we obtained this result for the particular case of $\kappa=1$. So for a symmetric and informationally complete POVM the set of probability distributions over its outcomes corresponding to quantum states is the same as the set of Bloch vectors, although somewhat shrunk.

Then given a quantum state \eqref{e:GenState} and a symmetric and informationally complete POVM $\cE\left(\kappa,\{\vec n_i\}_{i=1}^{d^2}\right)$ we can substitute its Bloch vector by the corresponding probability vector $\vec v$ using \eqref{e:v(b)SI-POVMS}
\begin{align}
  \rho &= \frac{1}{d}\I_d+\frac{1}{\kappa}\left(\sum_{i=1}^{d^2}p_i\vec t_i\right)\cdot\boldsymbol\sigma && \text{by \eqref{e:def v}}\notag\\
  &= \frac{1}{d}\I_d+\frac{d+1}{\kappa}\left(\sum_{i=1}^{d^2}p_i\vec n_i\right)\cdot\boldsymbol\sigma, && \text{by \eqref{e:t(n)}}\label{e:rho(prob)}
\end{align}
where $p_i$ is the probability of obtaining outcome $i$ when we measure $\cE\left(\kappa,\{\vec n_i\}_{i=1}^{d^2}\right)$ on $\rho$, note that $\kappa$ is not the purity index of $\rho$ but the corresponding parameter of the measured symmetric POVM.

\section{Conclusions}
We have used the spherical-coordinate method to describe symmetric POVMs not necessarily pure or informationally complete. The results were as follows. A symmetric POVM $\cE\left(\kappa,\{\vec n_i\}_{i=1}^N\right)$ for measuring a $d$-dimensional quantum system
\begin{itemize}
  \item cannot have less than $N_{min}(\kappa)$ outcomes, see equation \eqref{e:Nmin}.
  \item cannot have more than $d^2$ outcomes.
  \item exists irrespective of the orientation of the vectors $\{\vec n_i\}_{i=1}^N$, the dimension $d$ and $N\in\{2,3,\ldots,d^2\}$ if $\kappa\leq 1/(d-1)$.
\end{itemize}

Chosen a symmetric POVM $\cE\left(\kappa,\{\vec n_i\}_{i=1}^N\right)$ we can represent a quantum state by the probability distribution that it induces over the outcomes of that symmetric POVM. We have found that the set of such probability distributions, $\cQ_{N-1}^{(d)}$, is related to the set of Bloch vectors, $\cB^{(d)}$ as follows
\begin{equation}\label{e:Relation Qgen to B}
  \cQ_{N-1}^{(d)}=\kappa\cB^{(d)\|}
\end{equation}
where $\cB^{(d)\|}$ is the orthogonal projection of $\cB^{(d)}$ onto $\Span(\{\vec n_i\}_{i=1}^N)$. Only when $N=d^2$ the symmetric POVM is informationally complete in the sense that knowing the probability distribution that a state induces over its outcomes is sufficient to determine the corresponding Bloch vector, in this case
\begin{equation*}
  \cQ^{(d)}=\kappa\cB^{(d)}.
\end{equation*}
the set of quantum probability distributions is equal, up to scale, to the set of Bloch vectors. Therefore if we have an informationally complete and symmetric POVM, $\cE\left(\kappa,\{\vec n_i\}_{i=1}^{d^2}\right)=\{E_i\}_{i=1}^{d^2}$, we can write out any density matrix, $\rho\in\cD^{(d)}$, entirely in terms of the probabilities $p_i=\Tr(\rho E_i)$, the parameter $\kappa$ and the directional vectors of $\cE\left(\kappa,\{\vec n_i\}_{i=1}^{d^2}\right)$
\begin{equation*}
  \rho =\frac{1}{d}\I_d+\frac{d+1}{\kappa}\left(\sum_{i=1}^{d^2}p_i\vec n_i\right)\cdot\boldsymbol\sigma
\end{equation*}

\bibliographystyle{siam}
\bibliography{MyBib}

\begin{thebibliography}{1}

\bibitem{Appleby07}
{\sc D.~Appleby}, {\em Symmetric informationally complete measurements of
  arbitrary rank}, Optics and Spectroscopy, 103 (2007), pp.~416--428.
\newblock arXiv:quant-ph/0611260.

\bibitem{Bengtsson&Weis&Zyczkowski13}
{\sc I.~Bengtsson, S.~Weis, and K.~\.Zyczkowski}, {\em Geometry of the set of
  mixed quantum states: An apophatic approach}, in Geometric Methods in
  Physics, P.~Kielanowski, S.~T. Ali, A.~Odzijewicz, M.~Schlichenmaier, and
  T.~Voronov, eds., Trends in Mathematics, Springer Basel, 2013, pp.~175--197.
\newblock arXiv:1112.2347 [quant-ph].

\bibitem{Bengtsson&Zyczkowski06}
{\sc I.~Bengtsson and K.~\.{Z}yczkowski}, {\em Geometry of Quantum States},
  Cambridge University Press, 2006.

\bibitem{Gour13}
{\sc G.~Gour}, {\em Construction of all general symmetric informationally
  complete measurements}.
\newblock arXiv:1305.6545 [quant-ph], 2013.

\bibitem{Kimura03}
{\sc G.~Kimura}, {\em The bloch vector for n-level systems}, Physics Letters A,
  314 (2003), pp.~339 -- 349.
\newblock arXiv:quant-ph/0301152.

\bibitem{Kimura&Kossakowski05}
{\sc G.~Kimura and A.~Kossakowski}, {\em The bloch-vector space for n-level
  systems: the spherical-coordinate point of view}, Open Systems \& Information
  Dynamics, 12 (2005), pp.~207--229.
\newblock arXiv:quant-ph/0408014.

\bibitem{Rastegin13}
{\sc A.~E. Rastegin}, {\em Notes on general sic-povms}.
\newblock arXiv:1307.2334 [quant-ph], 2013.

\bibitem{Rosado11}
{\sc J.~I. Rosado}, {\em Representation of quantum states as points in a
  probability simplex associated to a sic-povm}, Foundations of Physics, 41
  (2011), pp.~1200--1213.
\newblock arXiv:1007.0715.

\end{thebibliography}

\end{document}